\documentclass{PoS}

\usepackage{wrapfig}
\usepackage{xspace}
\usepackage{amsmath}
\usepackage{rviewport}
\usepackage{cite}
\usepackage[hidelinks]{hyperref}
\usepackage{cleveref}
\crefname{figure}{Fig.}{Figs.}

\newcommand{\eV}{\ensuremath{\mbox{e\kern-0.1em V}}\xspace}
\newcommand{\GeV}{\ensuremath{\mbox{Ge\kern-0.1em V}}\xspace}
\newcommand{\MeV}{\ensuremath{\mbox{Me\kern-0.1em V}}\xspace}
\newcommand{\GeVc}{\ensuremath{\mbox{Ge\kern-0.1em V}\kern-0.1em/\kern-0.05em c}\xspace}
\newcommand{\GeVcc}{\ensuremath{\mbox{Ge\kern-0.1em V}\kern-0.1em/\kern-0.05em c^2}\xspace}
\newcommand{\AGeV}{\ensuremath{A\,\mbox{Ge\kern-0.1em V}}\xspace}
\newcommand{\AGeVc}{\ensuremath{A\,\mbox{Ge\kern-0.1em V}\kern-0.1em/\kern-0.05em c}\xspace}
\newcommand{\MeVc}{\ensuremath{\mbox{Me\kern-0.1em V}\kern-0.1em/\kern-0.05em c}\xspace}
\newcommand{\MeVcc}{\ensuremath{\mbox{Me\kern-0.1em V}\kern-0.1em/\kern-0.05em c^2}\xspace}

\newcommand{\pT}{\ensuremath{p_\text{T}}\xspace}

\newcommand{\xF}{\ensuremath{x_\text{F}}\xspace}

\newcommand{\EposLong}{{\scshape Epos\,1.99}\xspace}
\newcommand{\QGSJetLong}{{\scshape QGSJet\,II-04}\xspace}

\newcommand{\DPMJetLong}{{\scshape DPMJet\,3.06}\xspace}
\newcommand{\SibyllLong}{{\scshape Sibyll\,2.1}\xspace}
\newcommand{\SibyllNewLong}{{\scshape Sibyll\,2.3}\xspace}
\newcommand{\EposLHCLong}{{\scshape Epos\,LHC}\xspace}

\newcommand{\CernVM}{\textsc{Cern\-\kern-0.05emVM}\xspace}

\def \dedx{d$E$/d$x$\xspace}
\def \p{$p$\xspace}
\def \pions{$\pi^\pm$\xspace}
\def \kaons{K$^\pm$\xspace}
\def \proton{p\xspace}
\def \antiproton{$\bar{\text{p}}$\xspace}
\def \kaonstar{K$^{*0}$\xspace}
\def \rhozero{$\rho^{0}$\xspace}
\def \ncl{$N_{\mathrm{cl}}$\xspace}
\def \shine{NA61/SHINE\xspace}
\def \pipi{$\pi^+\pi^-$\xspace}

\newcommand{\NewSibyll}{{\scshape Sibyll\,2.3c}\xspace}

\title{Measurements of Hadron Production in Pion-Carbon Interactions with NA61/SHINE at the CERN SPS}

\ShortTitle{Hadron Production in Pion-Carbon Interactions (NA61/SHINE)}

\author{\speaker{Raul R. Prado}$\,^{1,2}$ for the NA61/SHINE
  Collaboration\thanks{http://shine.web.cern.ch/content/author-list}\\
  $^{1}$IKP, Karlsruhe Institute of Technology (KIT), Postfach 3640, D-76021 Karlsruhe, Germany \\
  $^{2}$Instituto de F\'isica de S\~ao Carlos, Universidade de S\~ao Paulo, Brazil\\
  E-mail: \email{raul.prado@usp.br}}

\abstract{NA61/SHINE is a fixed target experiment designed to study hadron-proton, hadron-nucleus and nucleus-nucleus interactions at the CERN Super-Proton-Synchrotron. In this proceeding we present results on spectra of identified hadrons produced in pion-carbon production interactions, which are of fundamental importance to improve the extensive air shower modeling, and hence the interpretation of ultra-high-energy-cosmic-rays measurements. In particular, our measurements of (anti)baryons and \rhozero production in pion-carbon interactions can contribute to improve the predictions of muon production by air shower simulations using hadronic interaction models. In this contribution we discuss the data analysis and the results from pion-carbon collisions recorded at beam momenta of 158 and 350 \GeVc. The preliminary spectra of \kaons and \proton(\antiproton) are shown, as well as a comparison to predictions of hadronic interaction models used in air shower simulations.  Additionally, we present final results on the production of \rhozero, $\omega$ and \kaonstar resonances.}

\FullConference{35th International Cosmic Ray Conference --- ICRC2017\\
		10--20 July, 2017\\
		Bexco, Busan, Korea}

\begin{document}

\section{Introduction}

Indirect measurements of high energy cosmic rays through Extensive Air Showers (EAS)
generally require air shower simulations to be interpreted. These simulations
are performed by Monte Carlo codes that make use of hadronic interaction
models to describe the nucleus-air and hadron-air collisions along the shower
development~\cite{Engel2011}. During the last decades, a number of EAS experiments
(like HiRes-MIA~\cite{Hires1999,Mia1994}, KASCADE-Grande~\cite{Navarra2004} and Pierre
Auger Observatory~\cite{Auger2015}) have reported results indicating that
 simulations do not predict satisfactorily the muon
production in air showers~\cite{HiresMia2000,Kascade2013,
  AugerMPD2014,AugerHas2015,TopDown2016}. This discrepancy
between data and simulations has strong implications
mainly on the inferences of the cosmic rays
mass composition~\cite{Kampert2012}.

The production of (anti)baryon and \rhozero along the air shower
development have been shown to be very important mechanisms for
the muon production~\cite{Pierog2008,Drescher2007,Ostapchenko2013}.
Therefore, a proper prediction of
these particles spectra by the hadronic interaction models
could contribute to reduce the discrepancy between air shower simulations
and data.
Although the current hadronic models are tuned by making use
of a large body of proton-nucleus data, there is a lack of
measurements of particle production in $\pi$-nucleus interactions,
which are the most abundant hadronic interaction occurring in an EAS.

To address the lack of suitable data for the tuning of hadronic
interaction models used in air shower simulations,
the \shine experiment~\cite{Shine2014} collected new data with
negatively charged pion beams at 158 and 350\,\GeVc on a thin carbon
target. A selection of results
on the hadron and meson productions of these collisions
are presented in this proceeding.

\section{The NA61/SHINE experiment}

\shine (SHINE = SPS Heavy Ion and Neutrino Experiment)
is a fixed target experiment at the CERN SPS designed to study
hadron production in nucleus-nucleus and hadron-nucleus
collisions. Its physics goals comprise a) the strong interaction
program, which investigate the properties of the onset of
deconfinement and search for the critical point of strongly
interacting matter, b) the neutrino program,
to measure precisely the hadron production important to calculate
the neutrinos and antineutrino fluxes in the T2K neutrino experiment~\cite{T2K2011},
and c) the cosmic rays program, focused on the measurements of the
hadron and meson production which are most relevant for the modeling
of extensive air showers. The full description of the \shine experiment
and its science program can be found in Ref.~\cite{Shine2014}

The \shine detector measures charged particles produced
by the collision of the beam particles with the target through
a set of five Time Projection Chambers (TPC). Since two of the TPCs
are placed in the magnetic field produced by superconducting
dipole magnets, the charge and the momenta of the particles
can be measured and the achieved resolution on \p is of order of
$\sigma(p)/p^2 = 10^{-4}$ (\GeVc)$^{-1}$. Additionally, the
energy loss by unit of length (\dedx) in the TPCs is used in this
work for particle identification.

A beam detector system composed of scintillation and Cherenkov counters is
placed upstream of the detector to identify and measure the
beam particles. The position of the beam is measured by
set of three beam position detectors, also placed upstream of
the target.

The data analyzed in this work are collisions of a $\pi^-$ secondary
beam with a thin (2 cm) carbon target. Two data sets were recorded in
2009 with beam energies of 158 and 350 \GeVc.

\section{\pions, \kaons, \proton and \antiproton spectra}

In NA61/SHINE charged particles can be identified
 via the track-by-track measurement of the
 truncated mean of the deposited energy registered
by individual pads in the TPCs.

After splitting the data into bins of total and transverse momentum
(\p and \pT), the measured \dedx distributions are fitted by a generic
\dedx model that combines the contribution of 5 particle types ($e$,
$\pi$, $K$, $p$ and deuterons). The \dedx distributions for individual
particles are described by the model as asymmetric Gaussians in which
the resolution $\sigma$ depends on the number of clusters
$N_{\mathrm{cl}}$ in each track as $\sigma \sim
1/\sqrt{N_{\mathrm{cl}}}$.  The particle yields are obtained then by
integrating the fitted distributions.

%\begin{wrapfigure}{r}{0.5\textwidth}
%  \centering
%  \includegraphics[width=0.5\textwidth]{figures/dedx.pdf}
%  \caption{Example of a fitted \dedx distribution. The upper panels
%    correspond to positive particles and the bottom panels to negative
%    ones.}
%  \label{fig:hadron:dedx}
%\end{wrapfigure}

The fits of the \dedx distributions provide a precise particle
identification for most of the phase space bins. However,
for a number of bins, mainly at the low momentum region, the average
energy deposit for different particle types can be very similar, which implies
a large overlap between the two \dedx distributions. For these bins,
the yields of the overlapping particles cannot be precisely determined
by the \dedx fit method and therefore they must be excluded from the
particle spectra. To determine which bins to exclude, a large set of
simulations of the \dedx distributions were produced, reproducing the
same \p, \pT and \ncl distributions and similar particle yields
observed in data. After performing the \dedx fit procedure to the
simulated data set, the standard deviation and the average bias of
the yields were computed and the bins with a relative standard
deviation larger than 25\% for a given particle were excluded from
the analysis. For the remaining bins the average biases were corrected
and the statistical uncertainties were computed by the standard
deviation of the simulated data fits.

The particle yields obtained from the \dedx distribution fits were
then corrected for detector effects (acceptance and selection
efficiency) and feed-down contributions from weak decays and
re-interactions in the target and downstream detector material.
Simulated data using three hadronic interaction models
(\EposLong~\cite{Pierog2008}, \QGSJetLong~\cite{Ostapchenko2010} and
\DPMJetLong~\cite{DPMJet2001}) was used to compute the correction
factors.  While the correction factor dependence on the specific
hadronic model was observed to be $< 2\%$ for the acceptance and
selection efficiency, the differences between the feed-down
contribution can reach up to $10\%$, depending on the \p and \pT bin
and particle type. These differences were accounted for in the
systematic uncertainties of the final spectra.

\begin{figure}
  \centering
  \includegraphics[clip, rviewport=0 0.2 1 1,width=0.95\textwidth]{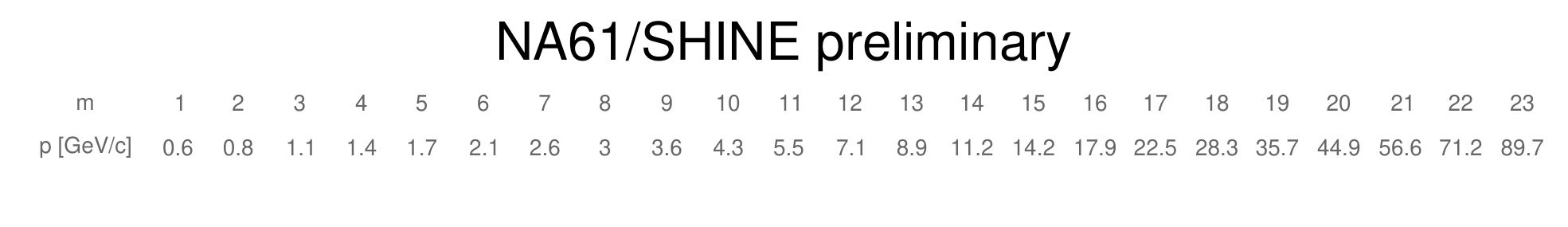}

  \includegraphics[clip, rviewport=0 0 0.92 0.91,height=8.5cm]{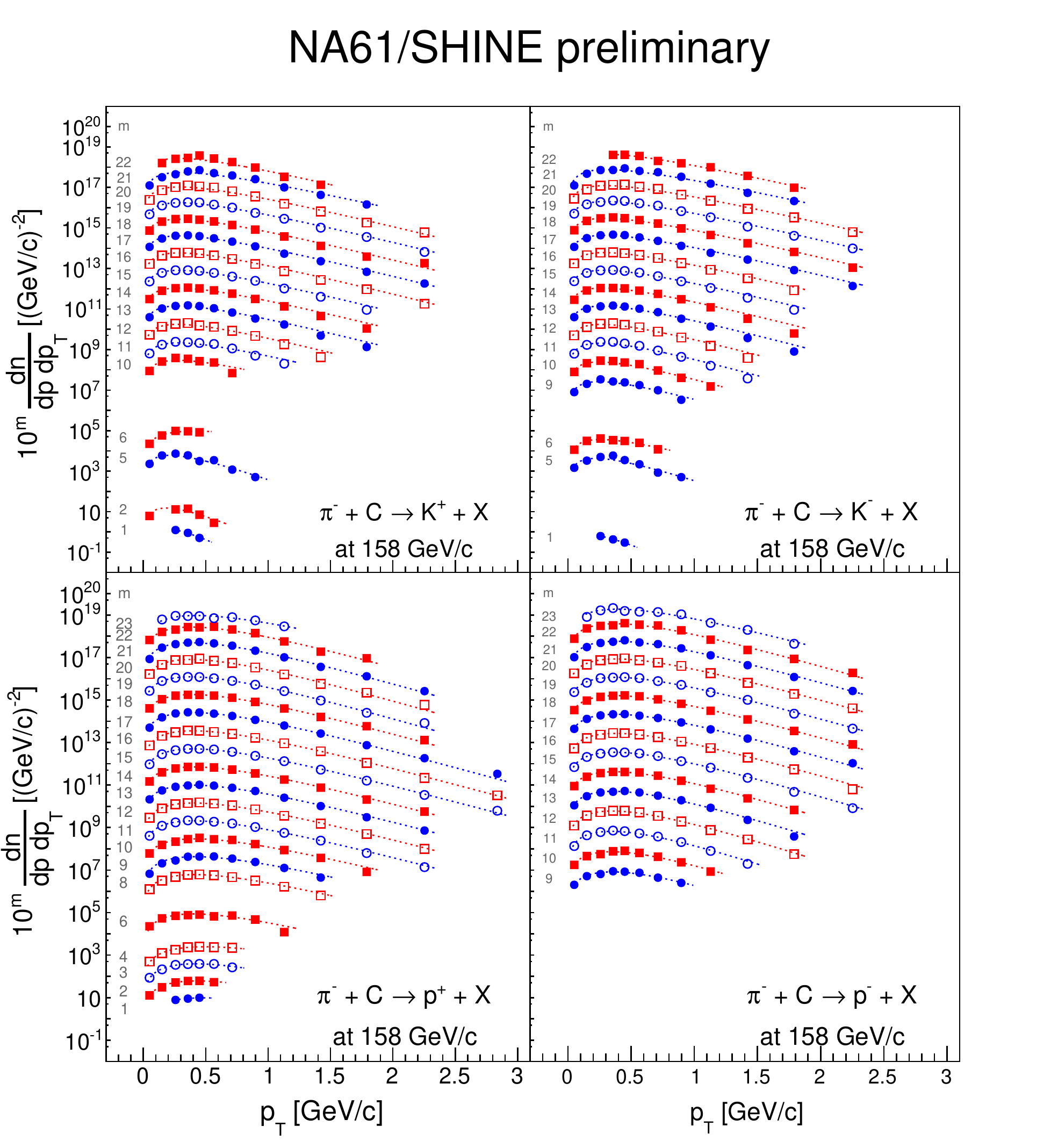}
  \includegraphics[clip, rviewport=0.1 0 0.92 0.91,height=8.5cm]{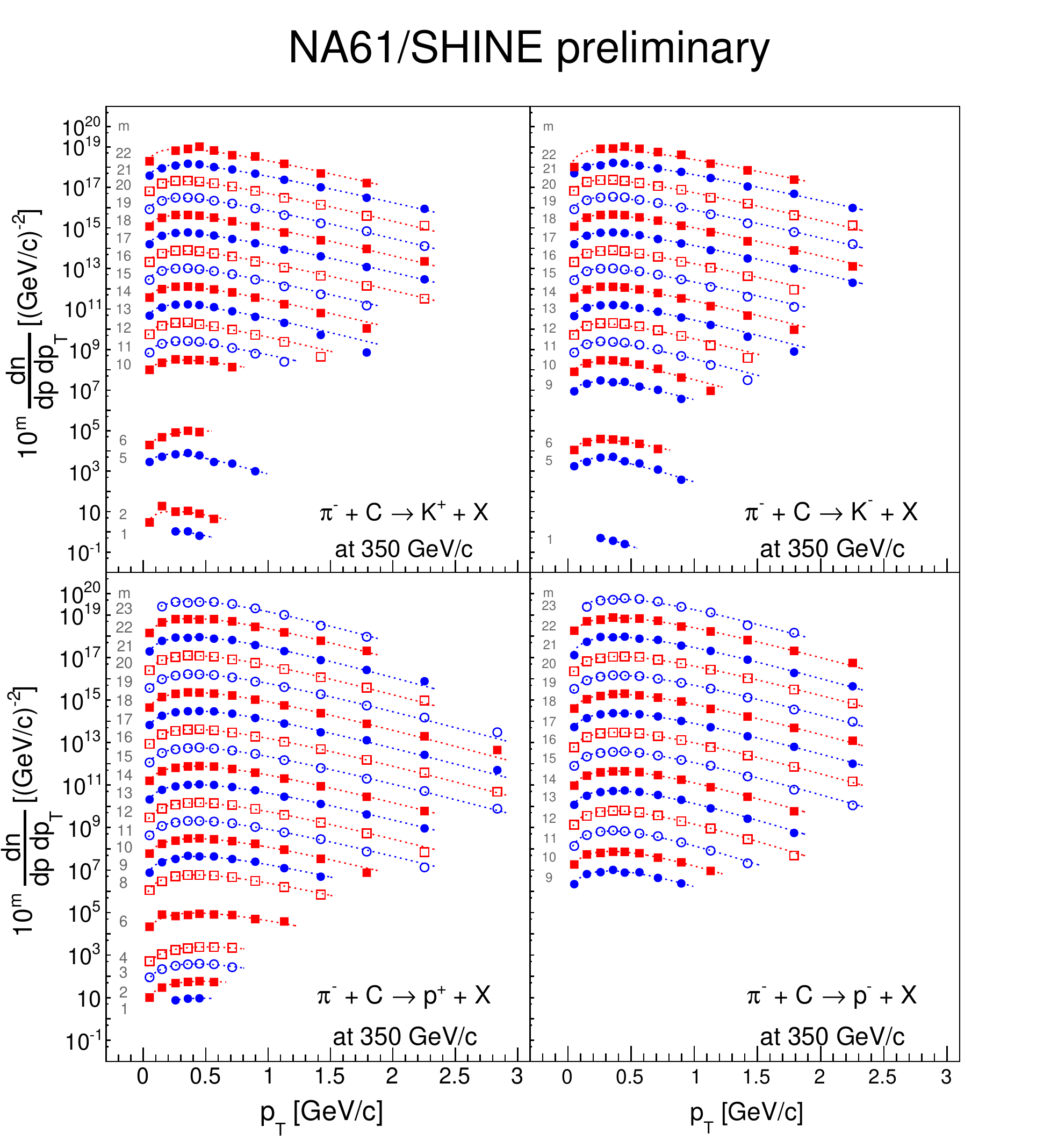}

  \caption{Spectra of \kaons and \proton(\antiproton) as function of \pT
    at 158 \GeVc (left) and 350 \GeVc (right). For each \p the spectrum
    was multiplied by $10^{m}$ with the value of $m$ shown on the left.}
  \label{fig:hadron}
\end{figure}

Finally, the average particle multiplicity, for each \p and \pT bin,
was computed by dividing the corrected yields by the total number of
target interactions, which in turn was obtained by correcting the
total number of triggered events by the trigger and event-selection
efficiency. This correction was also computed by using simulations
with the aforementioned hadronic interaction models and it amounts to
$92.5\pm3.5\%$ and $78.5\pm4.0\%$ for the 158 and 350 \GeVc data set,
respectively, where the uncertainties were estimated by the
differences between the factor obtained with different hadronic
models.  The final spectra of \kaons and \proton(\antiproton) are shown
in~\cref{fig:hadron}. The preliminary \pions spectra can be found in
Ref.~\cite{ICRC2015}.

For each \p bin, it was verified that the average multiplicity
as a function of \pT can be well described by an exponential
function in $m_{\text{T}}$, which means
$\frac{dn}{dp_{\text{T}}} \sim p_{\mathrm{T}} \, \exp[-(p_{\mathrm{T}}^2+m_0^2)^\frac{1}{2}/{T}]$,
with particle mass $m_0$ and inverse slope parameter $T$.
This function was fitted to the data and the result is shown
as dashed lines in~\cref{fig:hadron}.

The integrated spectra over \pT for
\pions, \kaons and \proton(\antiproton) are shown
in~\cref{fig:hadron:int158,fig:hadron:int350},
where they are compared to the predictions of
\EposLong, \DPMJetLong, \NewSibyll~\cite{Engel2017},
\QGSJetLong and \EposLHCLong~\cite{Pierog2013}.
Note that due to an improved subtraction of non-target interactions,
the updated \pions spectra shown here differ from the preliminary ones
presented in Ref.~\cite{ICRC2015} by $\leq 10\%$. 
To perform the integration, two functions were used to extrapolate
the \pT range which is not covered by the measured spectra:
an exponential in $m_{\text{T}}$ and a Gaussian convoluted with
an exponential. The average spectra obtained by using these functions are
shown in~\cref{fig:hadron:int158,fig:hadron:int350} and the difference
between both was added to the systematic uncertainties, being
in all cases $<1\%$.
The \p bins in which the contribution from the extrapolation
is larger than 5\% are not shown.

\def\figw{0.49}
\begin{figure}
  \centering
  \includegraphics[clip, rviewport=0 0.12 1 1,width=\figw\textwidth]{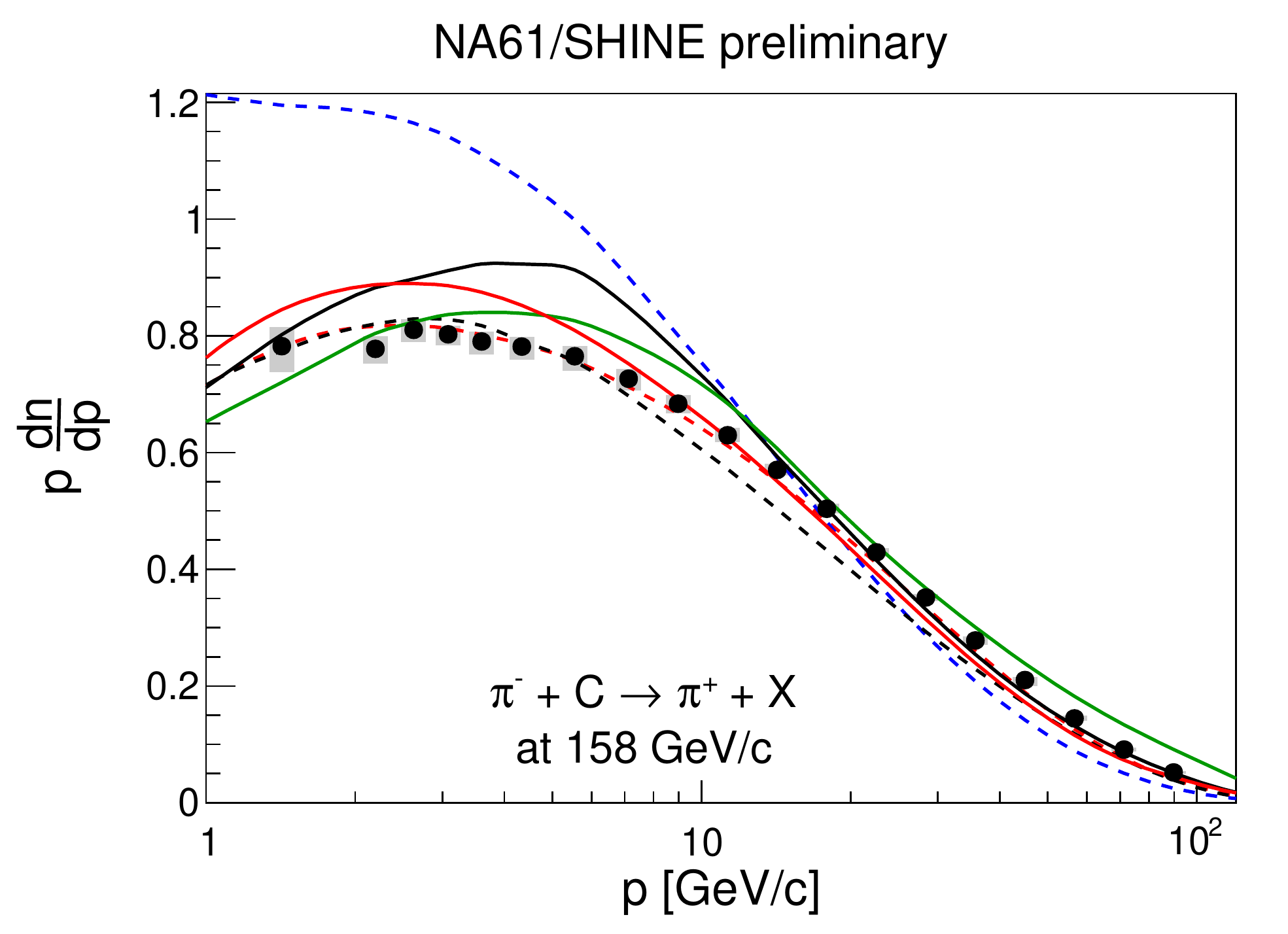}
  \includegraphics[clip, rviewport=0 0.12 1 1,width=\figw\textwidth]{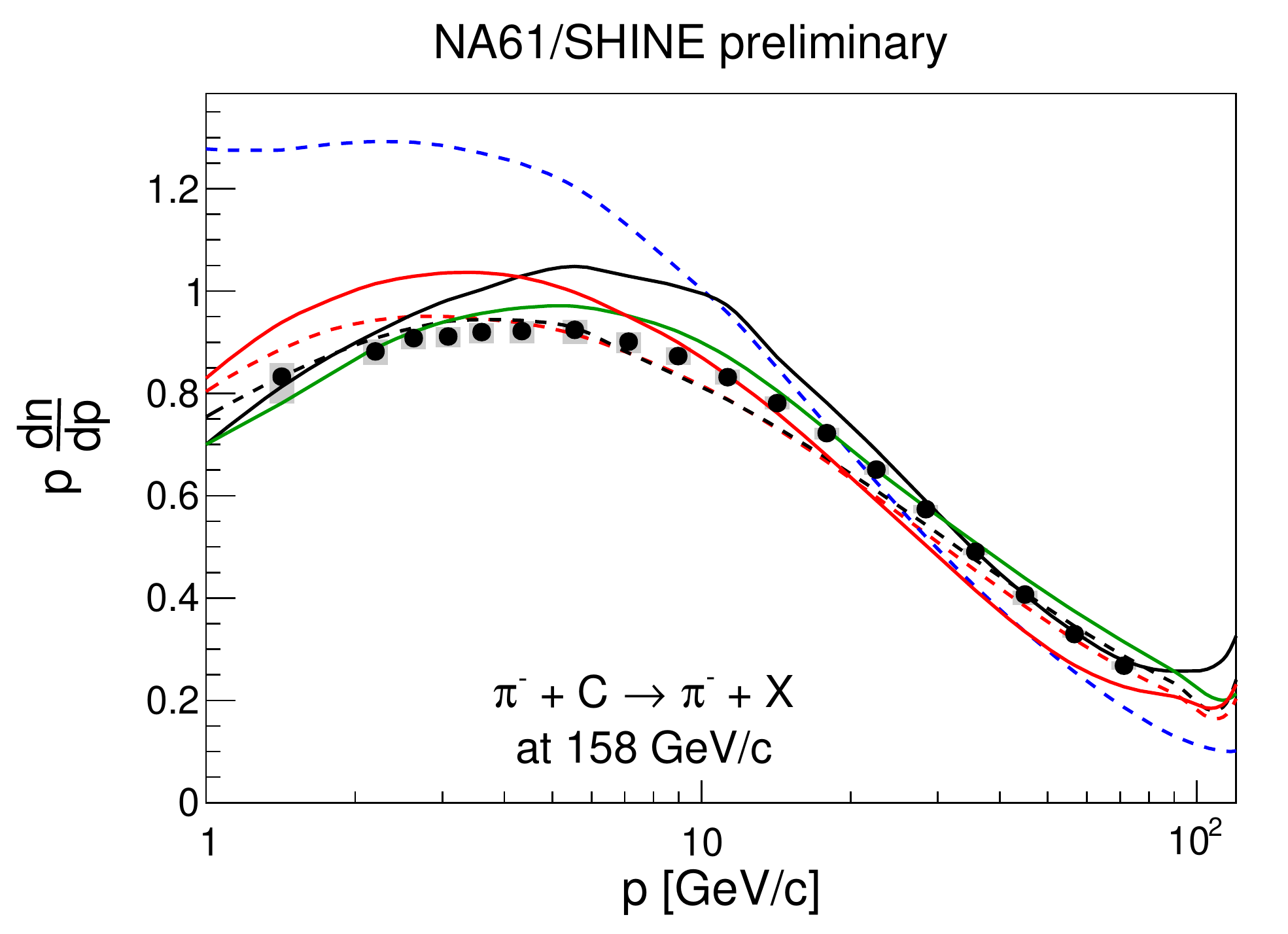}

  \includegraphics[clip, rviewport=0 0.12 1 0.9,width=\figw\textwidth]{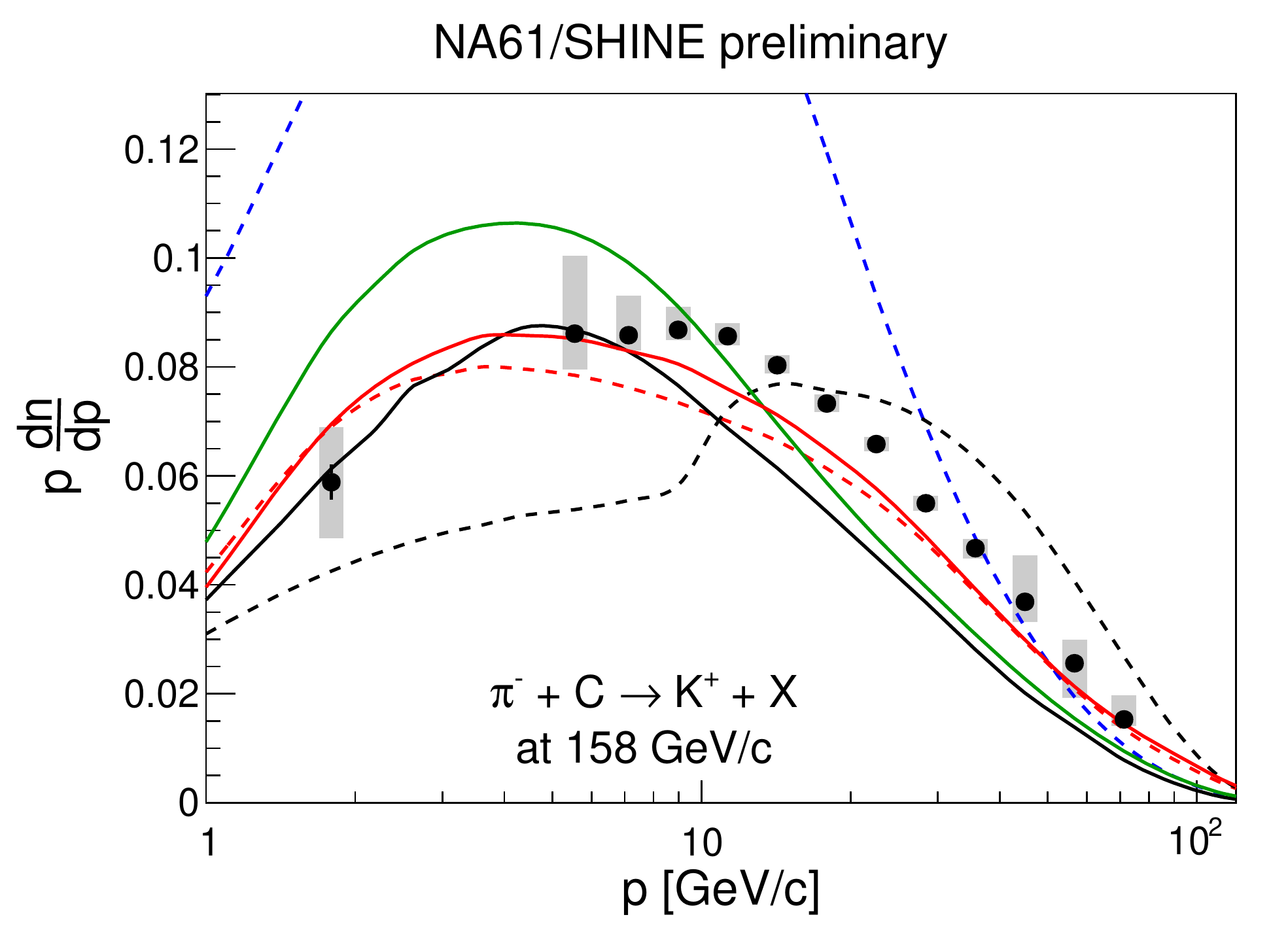}
  \includegraphics[clip, rviewport=0 0.12 1 0.9,width=\figw\textwidth]{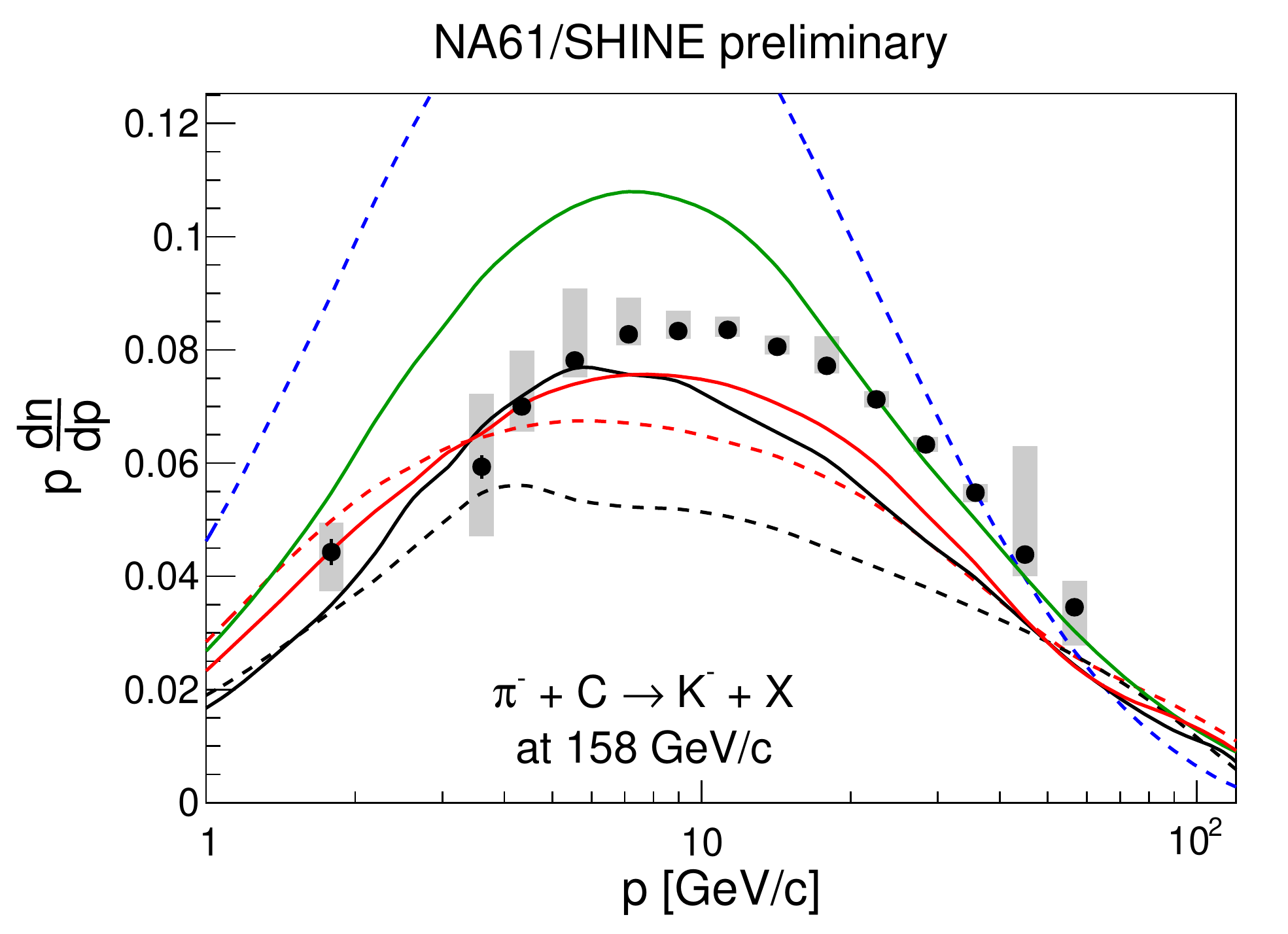}

  \includegraphics[clip, rviewport=0 0 1 0.9,width=\figw\textwidth]{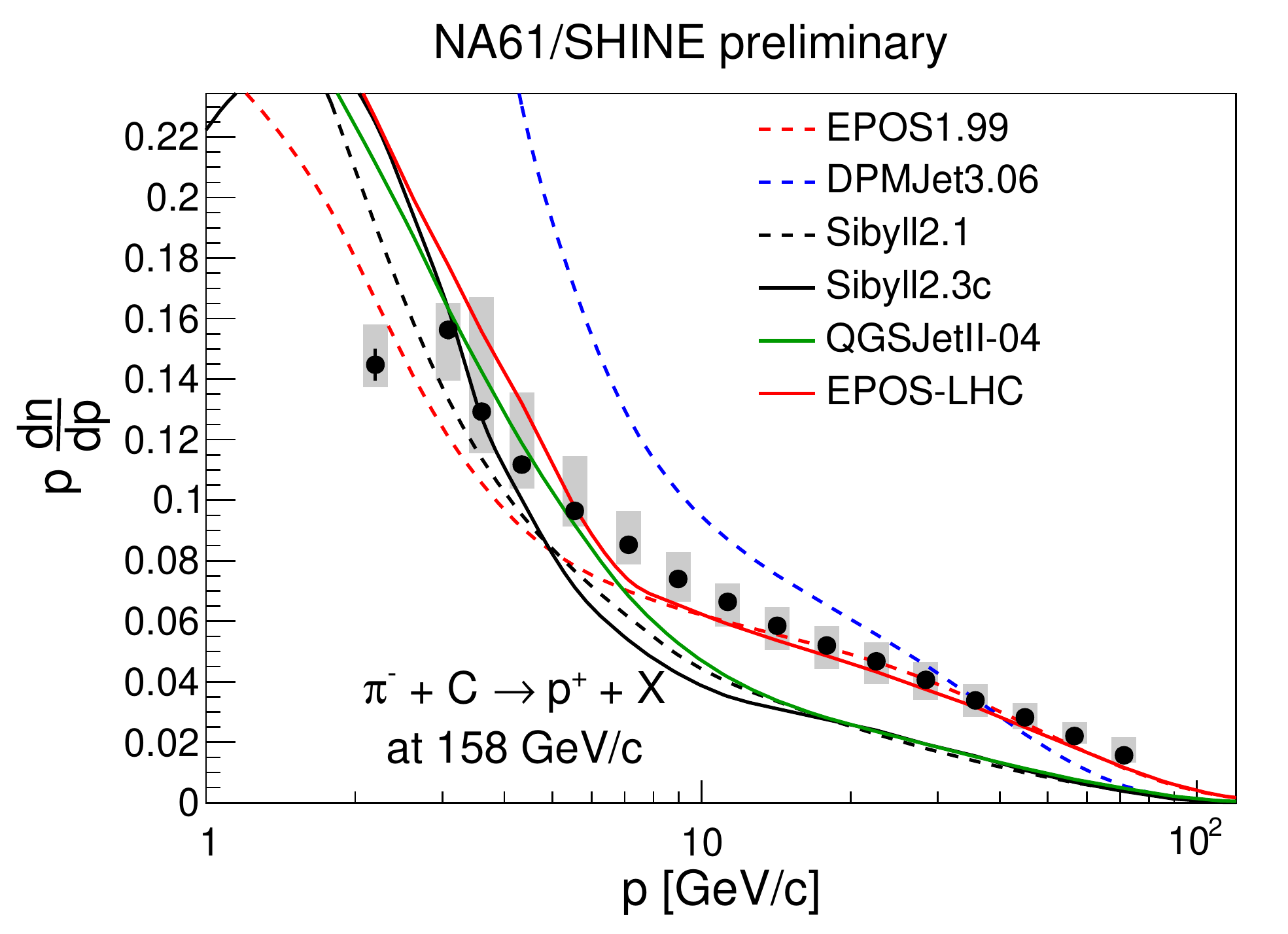}
  \includegraphics[clip, rviewport=0 0 1 0.9,width=\figw\textwidth]{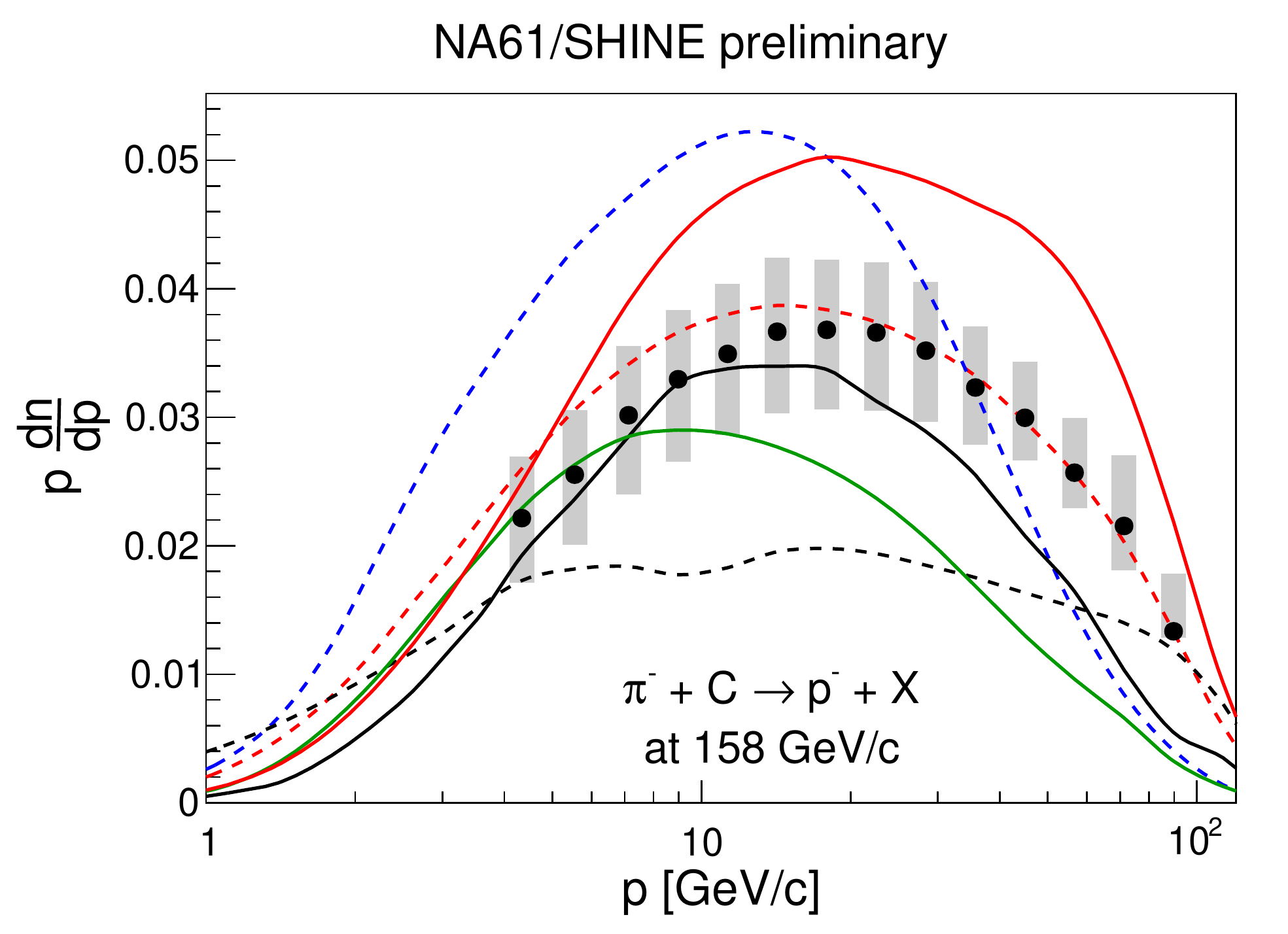}
  \caption{Spectra of \pions, \kaons and \proton(\antiproton) as a function of \p,
    integrated over \pT,
    for the 158 \GeVc data set. The statistical uncertainties are represented
    by bars and the systematic ones by gray bands.}
  \label{fig:hadron:int158}
\end{figure}

\begin{figure}
  \centering
  \includegraphics[clip, rviewport=0 0.12 1 1,width=\figw\textwidth]{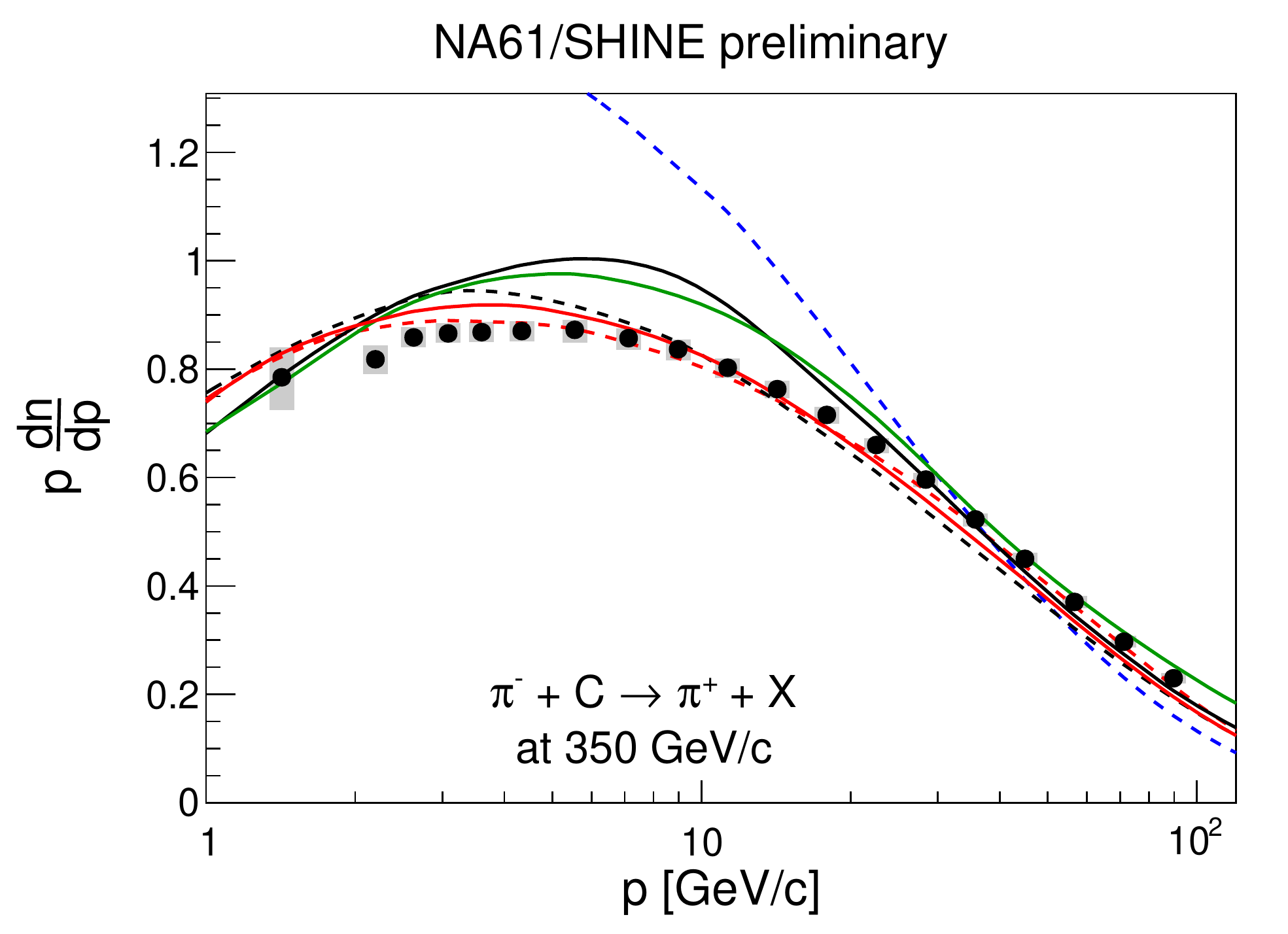}
  \includegraphics[clip, rviewport=0 0.12 1 1,width=\figw\textwidth]{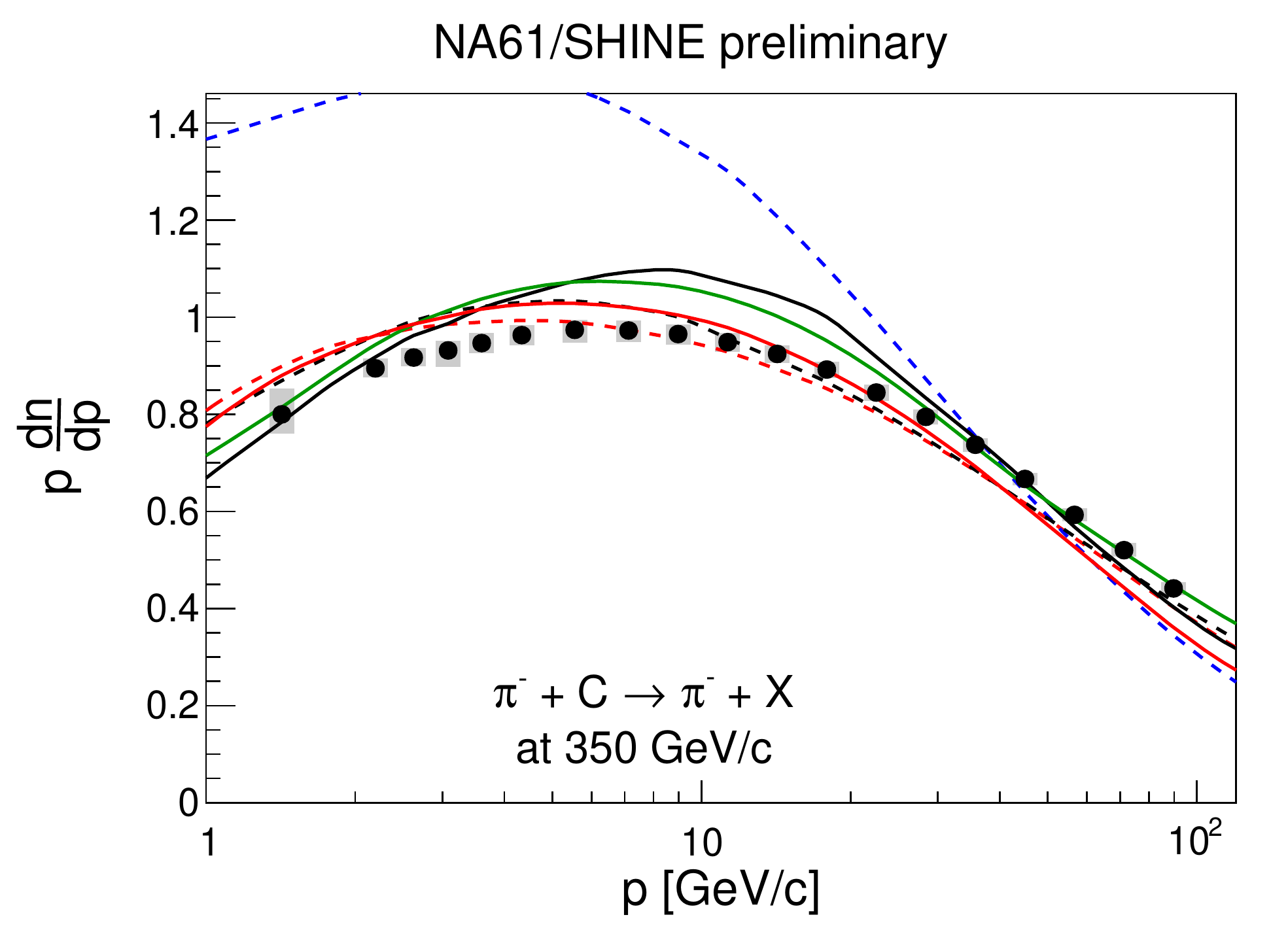}

  \includegraphics[clip, rviewport=0 0.12 1 0.9,width=\figw\textwidth]{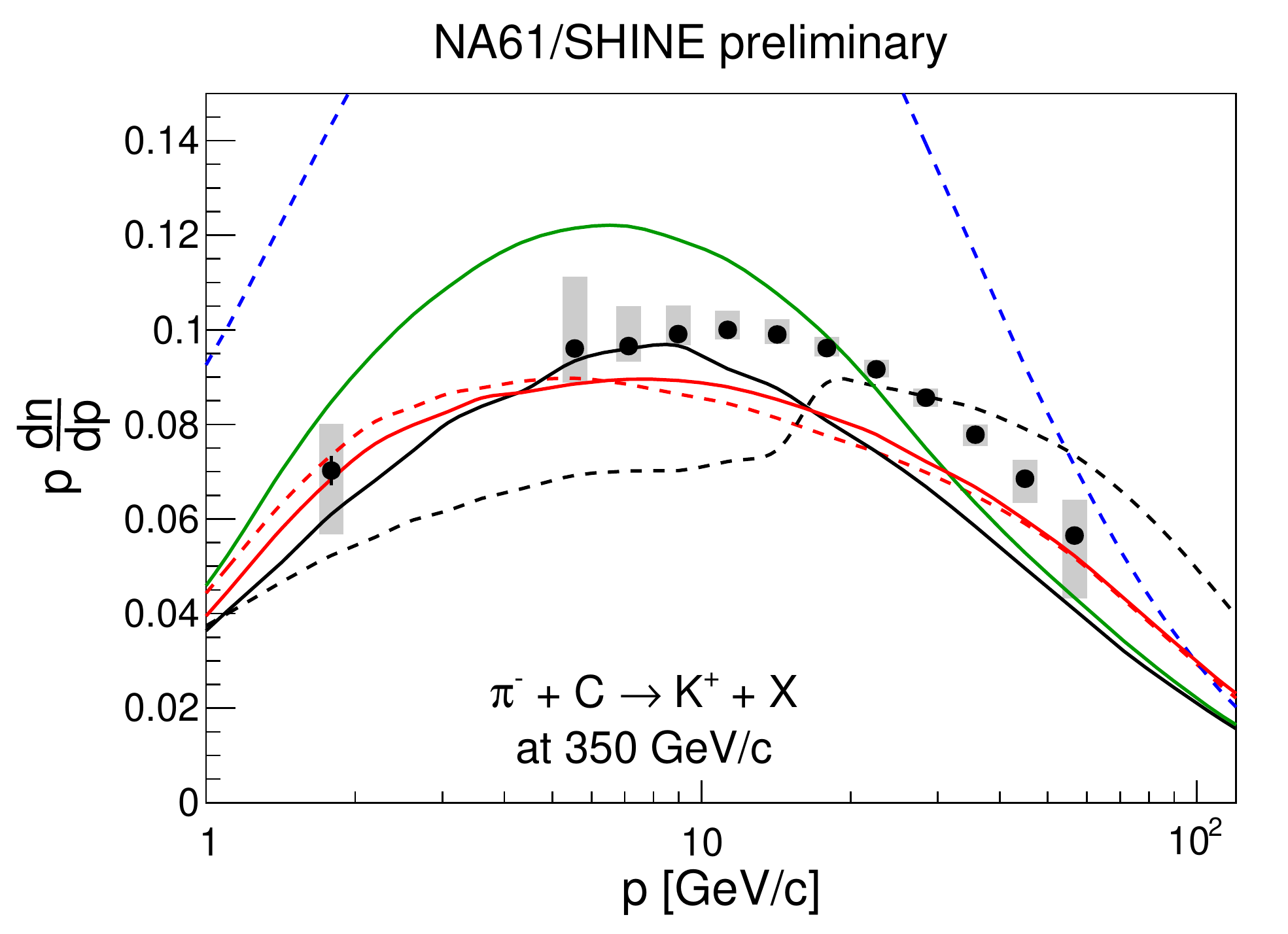}
  \includegraphics[clip, rviewport=0 0.12 1 0.9,width=\figw\textwidth]{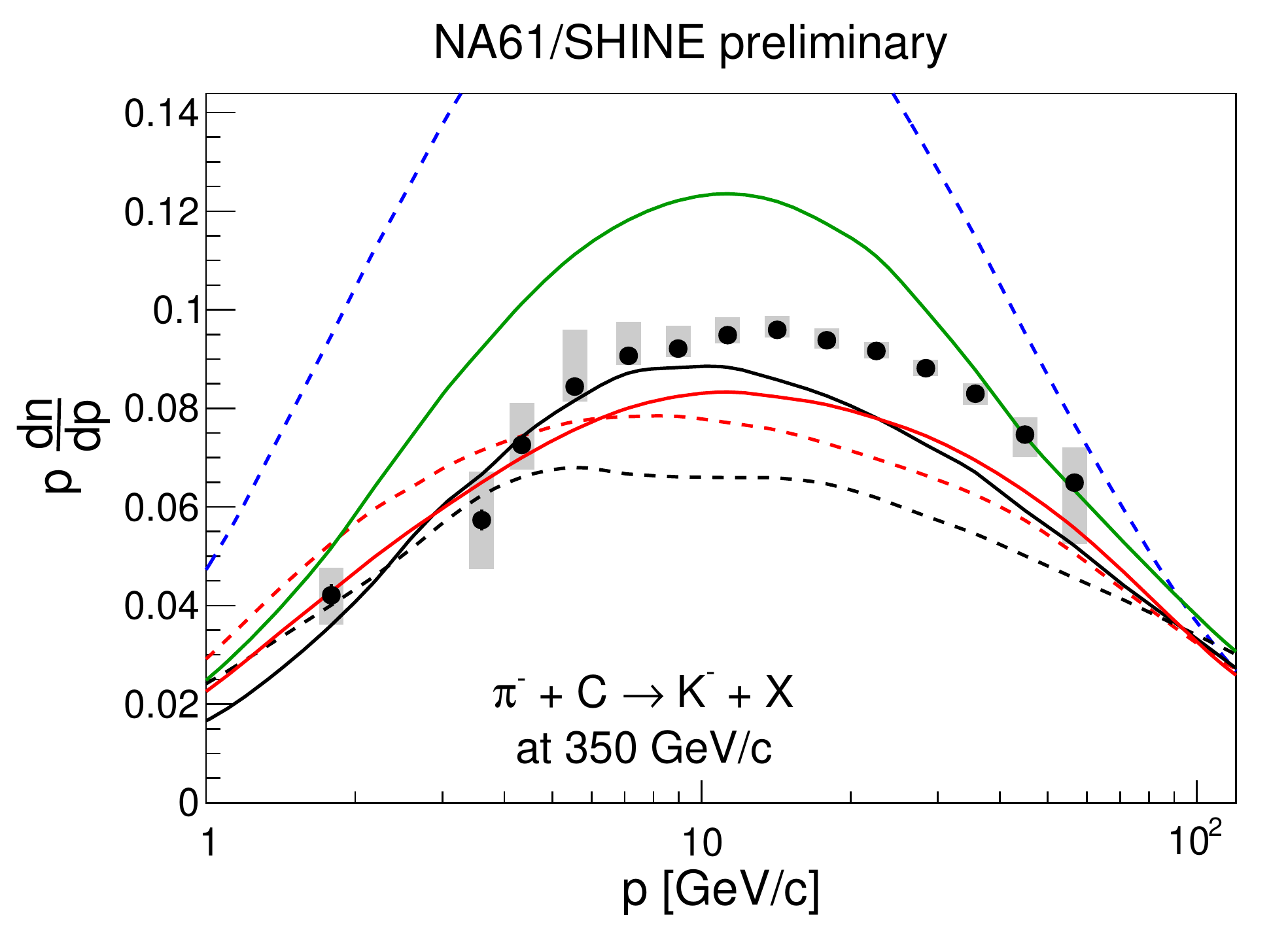}

  \includegraphics[clip, rviewport=0 0 1 0.9,width=\figw\textwidth]{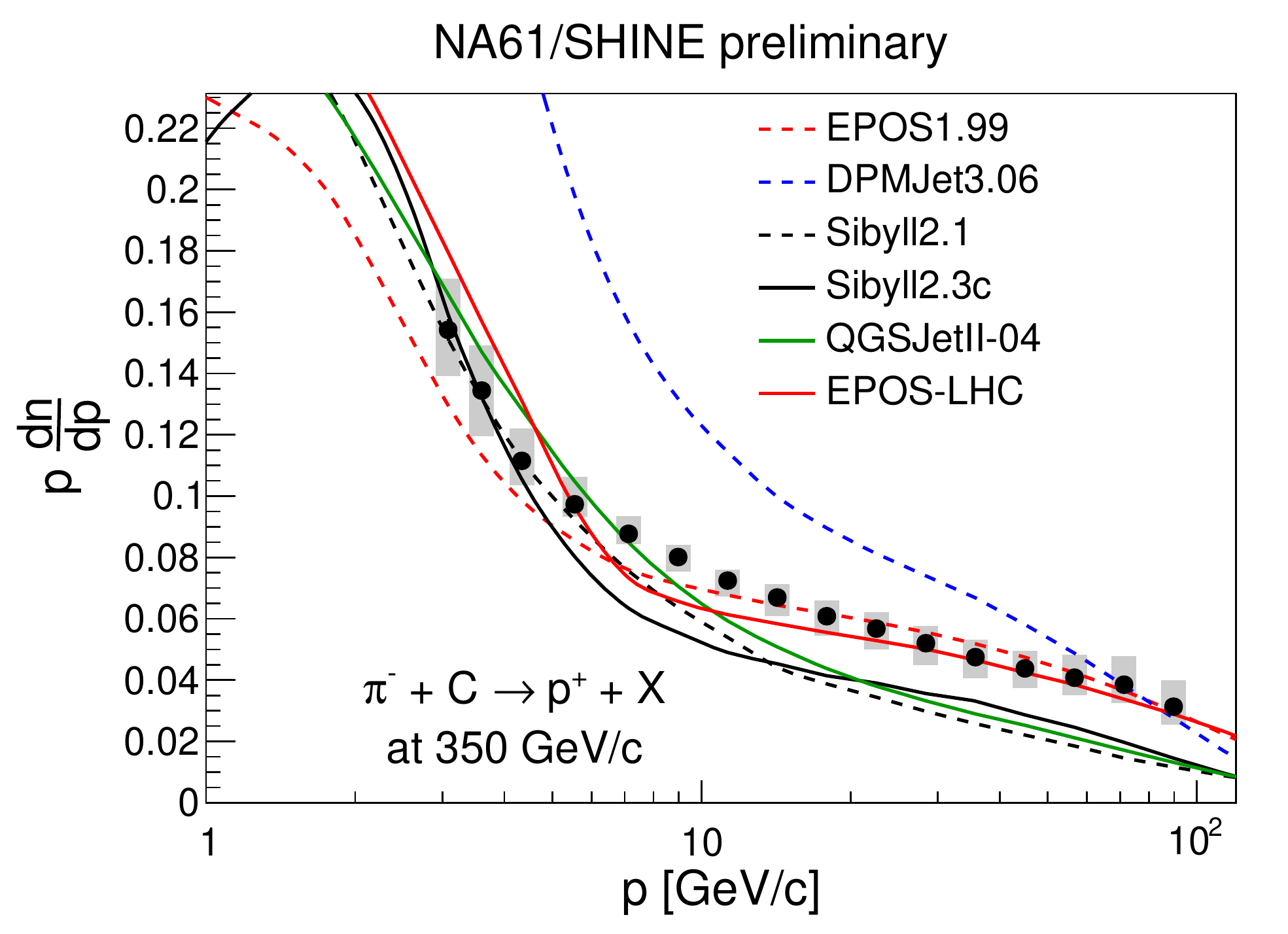}
  \includegraphics[clip, rviewport=0 0 1 0.9,width=\figw\textwidth]{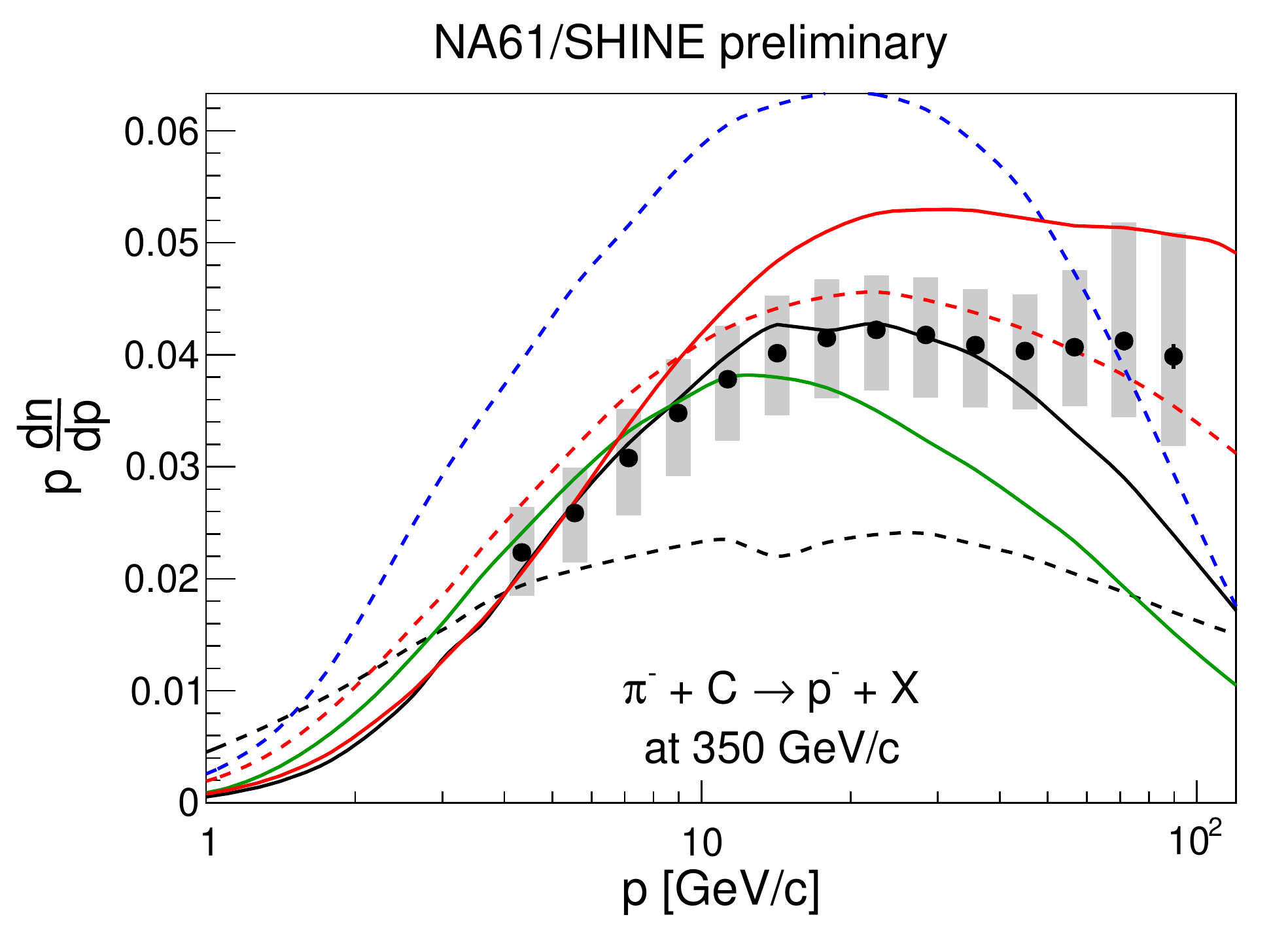}
  \caption{Spectra of \pions, \kaons and \proton(\antiproton) as a function of \p,
    integrated over \pT,
    for the 350 \GeVc data set. The statistical uncertainties are represented
    by bars and the systematic ones by gray bands. }
  \label{fig:hadron:int350}
\end{figure}

\section{\rhozero, $\omega$ and \kaonstar spectra}

The yields of the resonance mesons can be measured using the \shine
apparatus through the inclusive \pipi invariant mass spectra.
The resonance signal can be extracted by fitting templates of the \pipi
invariant mass
distribution, obtained from Monte Carlo simulation, to the measured
distribution. The Monte Carlo events were generated using \EposLong as hadronic
interaction model and they were passed through the full \shine
detector simulation and reconstruction chain.
To generate the template for the combinatorial background,
two approaches were applied:
charge mixing method, in which the $\pi^+\pi^+$ and $\pi^-\pi^-$ are
treated as the background, and Monte Carlo method, in which the
background mass distribution is obtained directly from
simulations.

The fitted yields were then corrected for the limited detector
acceptance, selection efficiency, detector trigger efficiency, fitting
bias and feed-down from re-interactions. Simulations using \EposLong
were used to compute the corresponding correction factors. Finally,
the average multiplicity was obtained by dividing the corrected yields
by the total number of target interactions. The complete analysis
description and results can be found in Ref.\cite{Resonance2017}.

\begin{figure}
  \centering
  \includegraphics[clip, rviewport=0 0.1 1 1,width=0.43\textwidth]{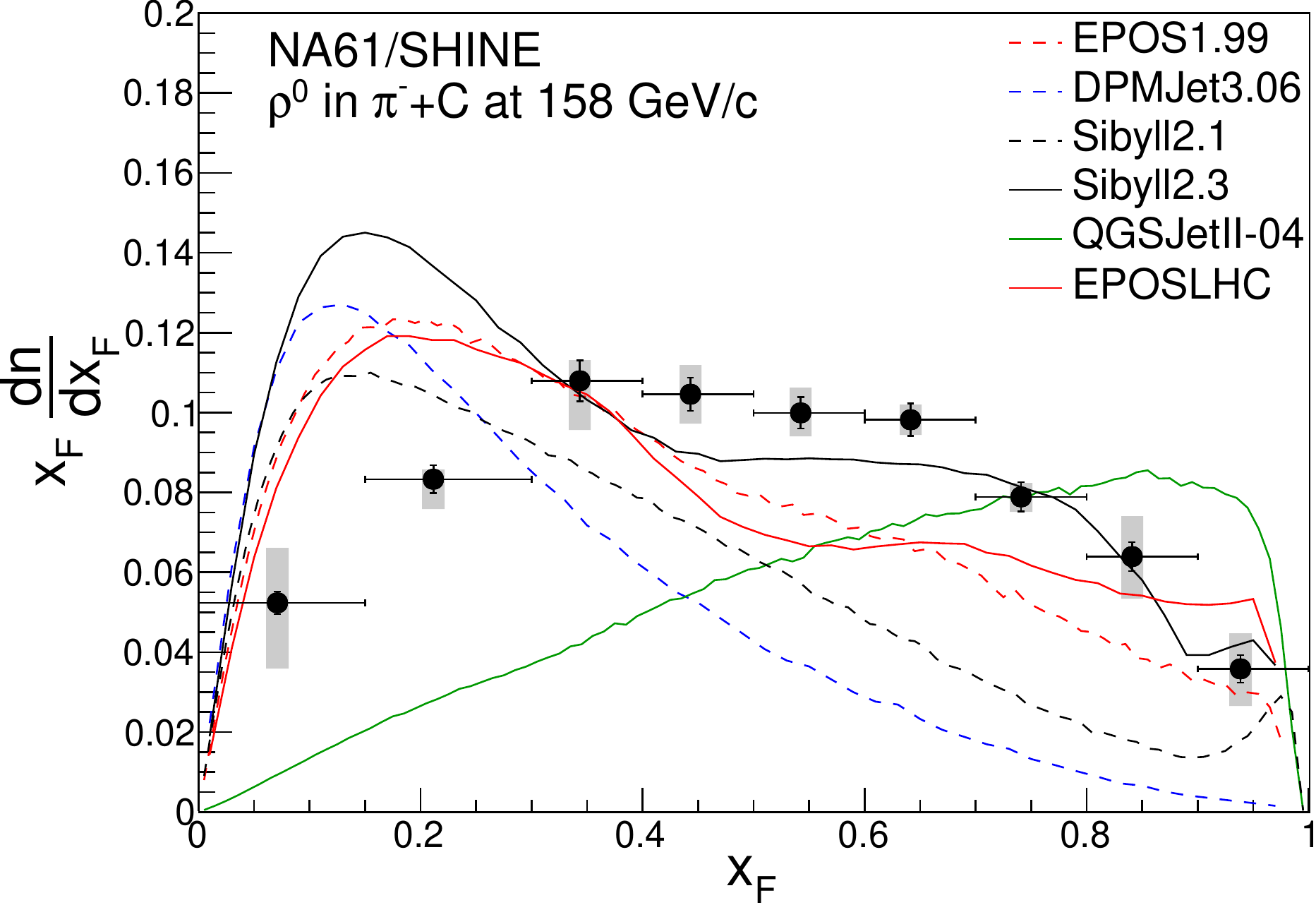}
  \includegraphics[clip, rviewport=0 0.1 1 1,width=0.43\textwidth]{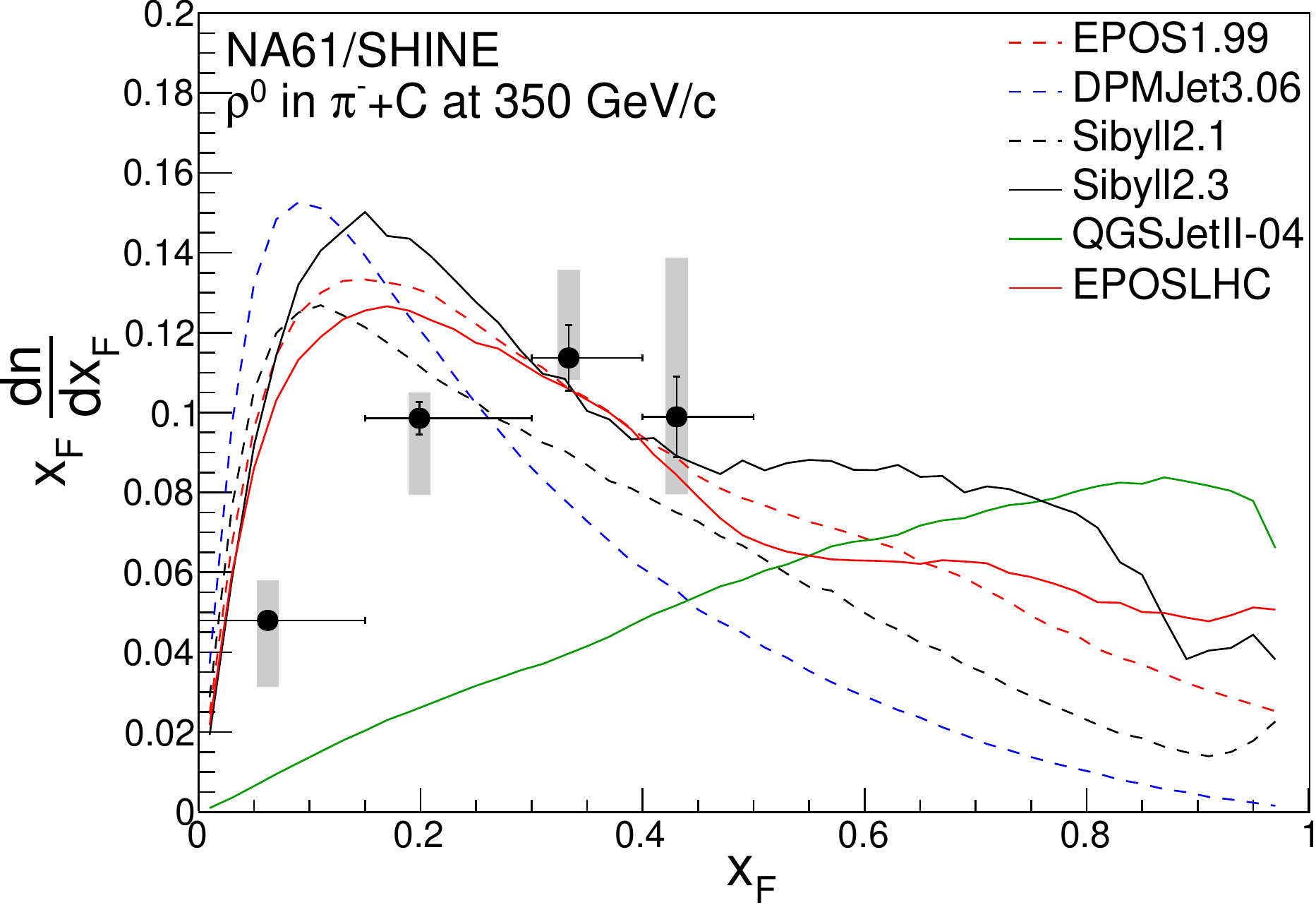}

  \includegraphics[width=0.43\textwidth]{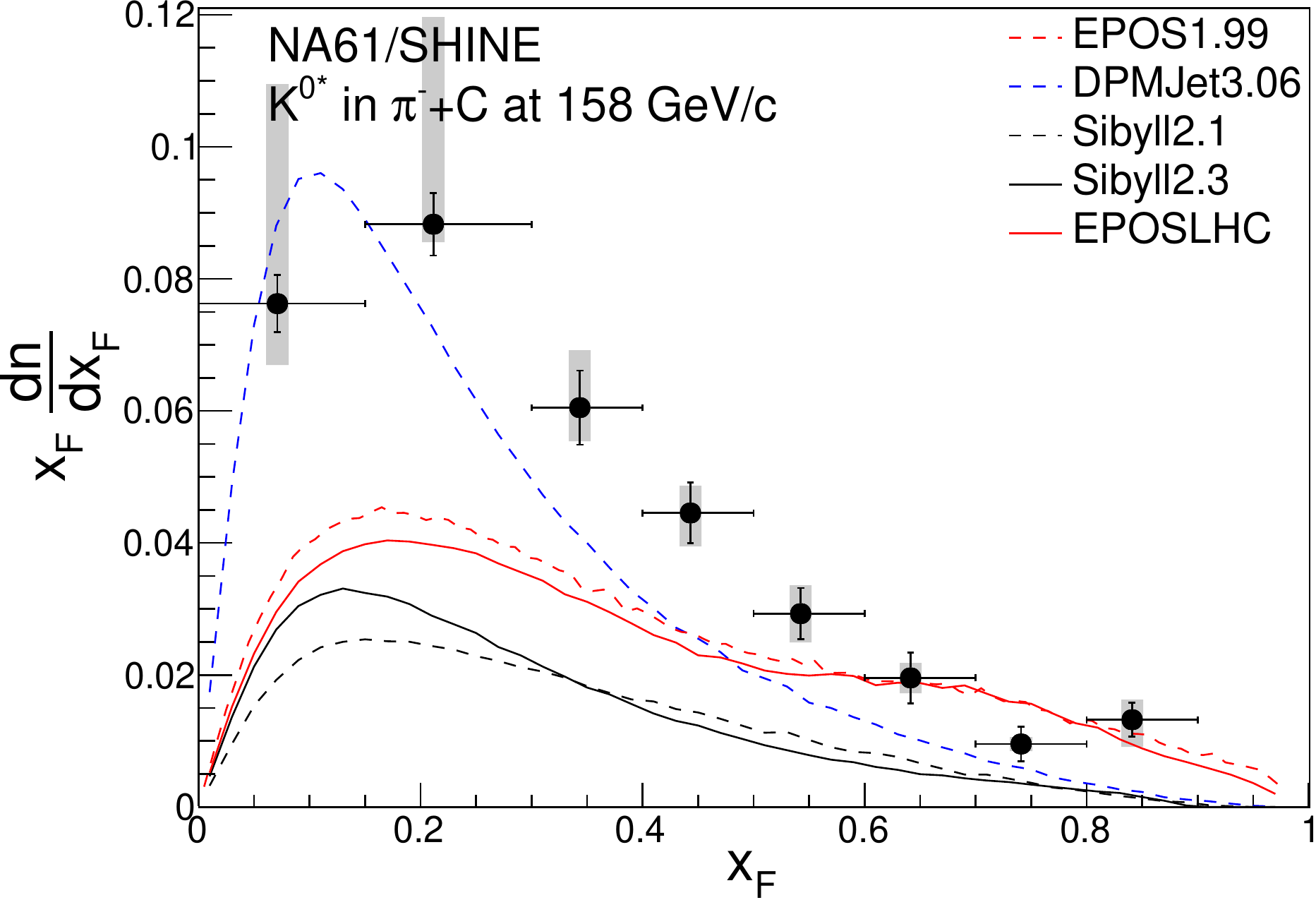}
  \includegraphics[width=0.43\textwidth]{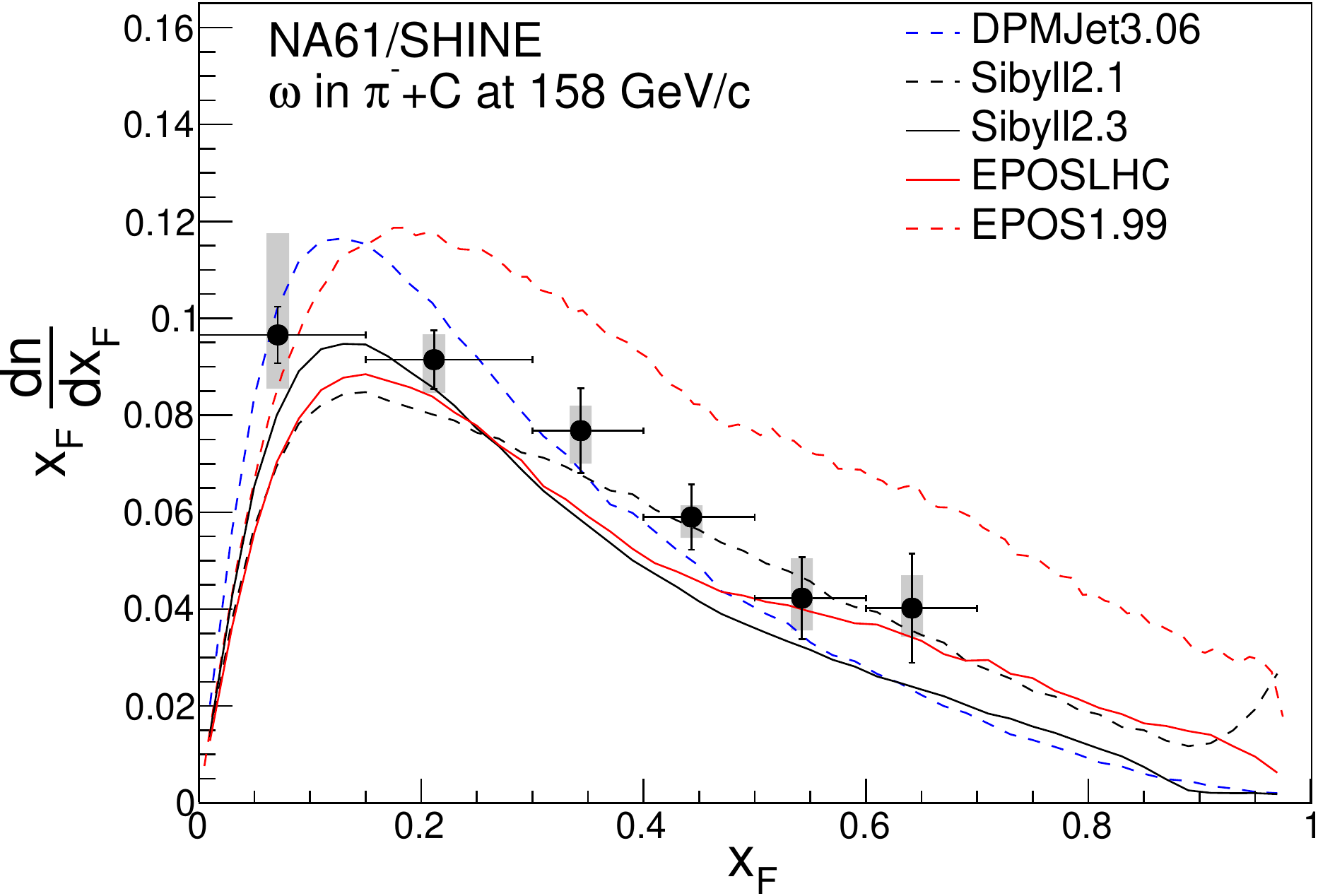}
  \caption{Spectra of \rhozero, $\omega$ and \kaonstar as a function of \xF.
    The statistical uncertainties are represented
    by bars and the systematic ones by gray bands~\cite{Resonance2017}.}
  \label{fig:resonance:spec}
\end{figure}

In~\cref{fig:resonance:spec}, the spectra of \rhozero, $\omega$ and \kaonstar
are shown and compared to the predictions of \EposLong, \DPMJetLong,
\SibyllLong, \SibyllNewLong~\cite{Riehn2015}, \QGSJetLong
and \EposLHCLong. While the \rhozero spectra are shown for beam energies of
158 and 350 \GeVc, the $\omega$ and \kaonstar spectra are limited to the 158 \GeVc
data set because of the large uncertainties obtained for the 350 \GeVc one.
Additionally, the \rhozero spectrum at 350 \GeVc is limited to \xF$< 0.5$
because of the limited acceptance of the detector at this beam energy.

The systematic uncertainties on the spectra were estimated by taking
into account the contributions of the difference between two methods
to determine the combinatorial background, differences on the
correction factors computed with different hadronic interaction models and
differences due to variations on the track and event selection criteria.

\section{Summary and conclusions}

Selected results of particle production in pion-carbon collisions
measured by \shine experiment were presented in this proceeding.
First we have shown the spectra of \pions, \kaons and \proton(\antiproton),
obtained by means of a particle identification analysis based on
the \dedx. Second, we have shown the spectra of \rhozero, $\omega$
and \kaonstar, obtained by means of a template fit method applied to
the invariant mass distribution of \pipi. The most relevant sources
of systematic uncertainties were estimated and presented with the final
spectra.

From the comparison of the measured \pions, \kaons and \proton(\antiproton)
spectra, one can see that none of the models provide a satisfactory description
of all particle production, at both energies. The special case of
\antiproton, which are important for air shower modeling,
\EposLong provides the better agreement, while its production is
overestimated by \EposLHCLong and \DPMJetLong, and  underestimated by
\QGSJetLong and \SibyllLong. Concerning the spectra of \rhozero,
an overall overestimation of its production by the models
is observed at \xF$<0.3$, while at \xF$>0.3$, an overall
underestimation is observed instead.

\section*{Acknowledgments}
{\small
We would like to thank the CERN EP, BE and EN Departments for the
strong support of NA61/SHINE.

This work was supported by the Hungarian Scientific Research Fund
(grants OTKA 68506 and 71989), the J\'anos Bolyai Research Scholarship
of the Hungarian Academy of Sciences, the Polish Ministry of Science
and Higher Education (grants 667\slash N-CERN\slash2010\slash0,
NN\,202\,48\,4339 and NN\,202\,23\,1837), the Polish National Center
for Science (grants~2011\slash03\slash N\slash ST2\slash03691,
2013\slash11\slash N\slash ST2\slash03879, 2014\slash13\slash N\slash
ST2\slash02565, 2014\slash14\slash E\slash ST2\slash00018 and
2015\slash18\slash M\slash ST2\slash00125, 2015\slash 19\slash N\slash ST2 \slash01689), the Foundation for Polish
Science --- MPD program, co-financed by the European Union within the
European Regional Development Fund, the Federal Agency of Education of
the Ministry of Education and Science of the Russian Federation (SPbSU
research grant 11.38.242.2015), the Russian Academy of Science and the
Russian Foundation for Basic Research (grants 08-02-00018, 09-02-00664
and 12-02-91503-CERN), , the National Research Nuclear
University MEPhI in the framework of the Russian Academic Excellence
Project (contract No. 02.a03.21.0005, 27.08.2013),
the Ministry of Education, Culture, Sports,
Science and Tech\-no\-lo\-gy, Japan, Grant-in-Aid for Sci\-en\-ti\-fic
Research (grants 18071005, 19034011, 19740162, 20740160 and 20039012),
the German Research Foundation (grant GA\,1480/2-2), the EU-funded
Marie Curie Outgoing Fellowship, Grant PIOF-GA-2013-624803, the
Bulgarian Nuclear Regulatory Agency and the Joint Institute for
Nuclear Research, Dubna (bilateral contract No. 4418-1-15\slash 17),
Bulgarian National Science Fund (grant DN08/11), Ministry of Education
and Science of the Republic of Serbia (grant OI171002), Swiss
Nationalfonds Foundation (grant 200020\-117913/1), ETH Research Grant
TH-01\,07-3 and the U.S.\ Department of Energy.

RRP would like to thank the financial support by
Funda\c{c}\~ao de Amparo \`a Pesquisa do Estado de S\~ao Paulo
(FAPESP, proc. 2016/12735-3).
}
\bibliographystyle{na61Utphys}
\bibliography{HadronProduction.bib}

\end{document}